\documentclass[pdflatex,sn-mathphys-num,iicol]{sn-jnl}

\usepackage{graphicx}
\usepackage{multirow}
\usepackage{amsmath,amssymb,amsfonts}
\usepackage{booktabs}
\usepackage{tabularx}
\usepackage{xspace}
\usepackage{url}
\usepackage{capt-of}
\usepackage{placeins}
\usepackage{balance}

\newcommand{\sys}{FinCacheServe\xspace}
\newcolumntype{Y}{>{\raggedright\arraybackslash}X}
\makeatletter
\gdef\orcidlogo{\includegraphics[height=1em]{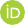}}
\makeatother

\begin{document}

\makeatletter
\def\au@and{\unskip,}
\makeatother

\title[\sys for Mutable RAG Serving]{\sys: Dependency-Consistent Answer Reuse for Cost-Efficient RAG Serving over Mutable Enterprise Documents}

\author*[1]{\fnm{Lingteng} \sur{Zeng}\,\orcid{https://orcid.org/0009-0006-0362-630X}}

\author[2]{\fnm{Yifan} \sur{Jin}\,\orcid{https://orcid.org/0009-0005-0032-5571}}

\affil*[1]{{\fontsize{9.5}{10.5}\selectfont \orgdiv{Faculty of Engineering}, \orgname{The Chinese University of Hong Kong}, \orgaddress{\city{Hong Kong SAR}, \country{China}}}}

\affil[2]{{\fontsize{9.5}{10.5}\selectfont \orgdiv{School of Computer Science and Technology}, \orgname{Beijing Institute of Technology}, \orgaddress{\city{Beijing}, \country{China}}}}

\abstract{Retrieval-augmented generation services over mutable enterprise documents repeatedly execute semantically equivalent analysis requests. Answer reuse can remove GPU-bound generation work, yet response caches require dependency consistency when filings, evidence chunks, and tool outputs change. \sys treats each generated answer as a serving object indexed by enterprise intent and guarded by document versions, evidence fingerprints, tool fingerprints, model identity, and decoding configuration. A vLLM implementation evaluates SEC-derived financial-document workloads with Qwen2.5 models. On a 2,230-request hosted 7B trace, \sys skips 53.27\% of LLM calls with zero observed dependency-stale outputs. Across three hosted 32B operator-suite seeds, it skips 53.31\% of 544 requests, compared with 38.97\% for versioned semantic caching and 22.43\% for grounded-style reuse. Capacity, backend, and SLO replays show oracle-bounded cache management, 100k-entry transactional metadata behavior, and 44.30\% lower estimated Wh per dependency-fresh 2s-SLO success than versioned semantic caching.}


\maketitle

\section{Introduction}\label{sec:introduction}

Cloud retrieval-augmented generation (RAG) services increasingly serve analytical workloads over corporate reports, filings, policies, tickets, and logs. A typical request retrieves evidence, builds an augmented prompt, calls a large language model (LLM), and returns an answer with cited context. This execution path spends GPU memory, prefill time, and decoding tokens even when the request repeats a previous analytical intent. The repeated work becomes significant in dashboards, compliance workflows, financial analysis assistants, and agentic systems that issue recurring questions over slowly changing corpora.

Caching is the natural performance mechanism for repeated serving work. LLM-serving systems have reduced GPU pressure through KV-cache management, prefix reuse, prefill reuse, and context scheduling \cite{kwon2023vllm,zheng2023sglang,jin2024ragcache,yao2024cacheblend,jiang2025contextpilot,feng2025adaptcache}. Those mechanisms accelerate execution after the request reaches the model. Answer reuse attacks a different bottleneck. A valid answer-cache hit can bypass the model invocation, removing both prefill and decoding for that request.

Mutable enterprise documents complicate answer reuse. A semantically similar query can require a fresh answer when a filing amendment changes a source table, when an extraction pipeline refreshes evidence chunks, when a ratio computation changes, or when serving moves to a different model/configuration. A cache hit based on text similarity or a fixed time-to-live window can therefore expose stale evidence. Financial filings make this dependency problem concrete and reproducible: each answer is tied to a company, period, filing scope, evidence set, tool output, and filing version. The workload contains repeated analytical intent, yet the validity of reuse depends on a fine-grained dependency contract.

\sys is a serving-layer system for dependency-consistent answer reuse. The system caches generated answers as materialized serving objects. Each entry records normalized financial identity, evidence fingerprints, tool-output fingerprints, source-document versions, model identity, and decoding configuration. The reuse gate admits an answer when the incoming request matches the recorded task identity and every recorded dependency remains compatible with the current metadata plane. Filing updates propagate through a document-to-answer reverse dependency index before later cache reads observe the updated version store.

This design targets a supercomputing-relevant performance point: reducing GPU-bound LLM invocations in hosted RAG serving while keeping a precise consistency contract for mutable evidence. The implementation integrates an answer cache, a retrieval cache, a FAISS-backed retriever \cite{johnson2021faiss}, and vLLM-hosted Qwen2.5 models \cite{kwon2023vllm,yang2024qwen25}. The evaluation uses SEC-derived workloads as a public high-value enterprise-document benchmark, with hosted 7B/14B/32B runs, strong answer-cache baselines, request-level provider-call accounting, capacity-limited cache-management replay, online policy sensitivity, metadata-backend validation, energy sensitivity, and interleaved query/update stress.

The paper makes four contributions.

\begin{enumerate}
  \item \textbf{Dependency-consistent answer reuse.} \sys defines an answer-cache contract that binds reuse to financial intent, source-document versions, evidence fingerprints, tool-output fingerprints, model identity, and decoding configuration.
  \item \textbf{GPU-serving implementation.} The implementation integrates answer-level reuse with a vLLM-backed RAG serving path and records LLM invocations, skipped calls, token estimates, freshness events, and stale-origin categories at request granularity.
  \item \textbf{Performance and safety evaluation.} A 2,230-request hosted 7B trace shows 53.27\% skipped LLM calls with zero observed dependency-stale outputs. Three hosted 32B operator-suite seeds show 53.31\% skipped calls, exceeding versioned semantic, grounded-style, entity-period semantic, and exact-invalidation baselines under the same dependency-stale criterion.
  \item \textbf{Metadata-plane validation.} Field-level ablations, capacity-limited admission and eviction replay, policy-history sensitivity, adversarial near-collision probes, SLO and energy replay, human metadata audit, reverse-index scaling, transactional-backend validation, and interleaved query/update stress characterize the mechanism behind the performance gain and the consistency behavior of the metadata path.
\end{enumerate}

\section{Related work}\label{sec:related}

\subsection{RAG over financial documents}

Financial question answering requires evidence selection, numerical reasoning, and period-aware interpretation. FinQA studies numerical reasoning over financial reports \cite{chen2021finqa}, and FinanceBench evaluates financial question answering against professional-style evidence requirements \cite{islam2023financebench}. SEC EDGAR provides public APIs for filing metadata and submissions \cite{sec2025edgar}. These sources motivate a reproducible mutable-document benchmark for RAG serving: requests are repetitive, answers are valuable, and source versions matter.

\subsection{LLM serving caches}

vLLM introduced PagedAttention to improve KV-cache memory management for high-throughput LLM serving \cite{kwon2023vllm}. SGLang exploits prefix sharing in structured language-model programs \cite{zheng2023sglang}. RAGCache and CacheBlend reuse knowledge or cached states in RAG execution \cite{jin2024ragcache,yao2024cacheblend}. ContextPilot and AdaptCache further optimize long-context or KV-cache reuse \cite{jiang2025contextpilot,feng2025adaptcache}. These systems reduce the cost of model execution. \sys complements them by moving reuse to the final answer layer, where a valid hit removes the model invocation from the request path.

\subsection{Semantic response caching}

Semantic caching reuses responses for similar inputs and can reduce LLM serving cost \cite{liu2025semanticcache}. Krites adds asynchronous verification for tiered semantic-cache promotion \cite{singh2026krites}. GroundedCache-style routing adds evidence overlap and source-validity gates to RAG answer reuse \cite{shah2026groundedcache}. MinCache uses hierarchical exact, resemblance, and semantic matching for efficient chatbot caching \cite{haqiq2025mincache}. \sys adds a domain-specific dependency contract for mutable enterprise documents: answer reuse depends on financial identity, period, document scope, evidence fingerprints, tool outputs, model identity, generation parameters, and filing versions.

\subsection{Cache consistency and provenance}

HTTP caching separates reuse from validation by requiring freshness metadata and revalidation semantics \cite{nottingham2022rfc9111}. Materialized-view maintenance propagates base-data updates to derived results \cite{gupta1995materialized,mistry2001materialized}. Data provenance records why and where derived data were produced \cite{buneman2001whywhere}. Large-scale data serving systems expose the performance-consistency tradeoff under updates \cite{decandia2007dynamo,cooper2008pnuts}. \sys applies these principles to RAG serving, where the cached object is a natural-language answer backed by mutable evidence and serving-time tools.

\section{System model and consistency target}\label{sec:model}

The service receives a request $r=(q,m,\theta)$, where $q$ is the user query, $m$ is the model identifier, and $\theta$ is the decoding configuration. Retrieval returns evidence chunks $E_r$ from documents $D_r$. Each document $d$ has a version $v(d)$ in the metadata plane. Tool execution produces structured outputs $T_r$ such as ratio values, table slices, or extracted filing facts.

An answer-cache entry $c$ contains an answer $a$, a normalized request signature $s(c)$, source-document versions $V(c)$, evidence fingerprint $h_E(c)$, tool fingerprint $h_T(c)$, model identity $m(c)$, decoding configuration $\theta(c)$, and reverse-index memberships from source documents to answer entries.

The primary performance target is LLM skip rate:
\begin{equation}
\mathrm{skip\ rate}=\frac{\mathrm{requests\ served\ from\ answer\ cache}}{\mathrm{measured\ requests}}.
\end{equation}

The primary consistency target is dependency freshness. A served output is dependency-stale when the answer or prompt uses a cached dependency whose recorded source-document version, evidence fingerprint, or tool fingerprint conflicts with the current metadata state for the request. This definition covers stale answer-cache hits and stale retrieval-evidence reuse. It evaluates consistency with recorded dependencies; factual answer accuracy is evaluated separately through answer audit and source review.

The reuse gate has a linearization point $\tau$. A cached answer can be served at $\tau$ when all predicates below hold:
\begin{align}
s(r) &\simeq s(c), \\
\forall d\in D_c,\quad v_\tau(d) &= V(c,d), \\
h_E(r) &= h_E(c), \\
h_T(r) &= h_T(c), \\
m &= m(c), \quad \theta = \theta(c).
\end{align}
Filing updates linearize by updating the version store and invalidating dependent answer entries before later reuse-gate reads. Concurrent queries and updates are ordered by this metadata-plane linearization.

\section{\sys design}\label{sec:design}

\begin{figure*}[t]
  \centering
  \includegraphics[width=\textwidth]{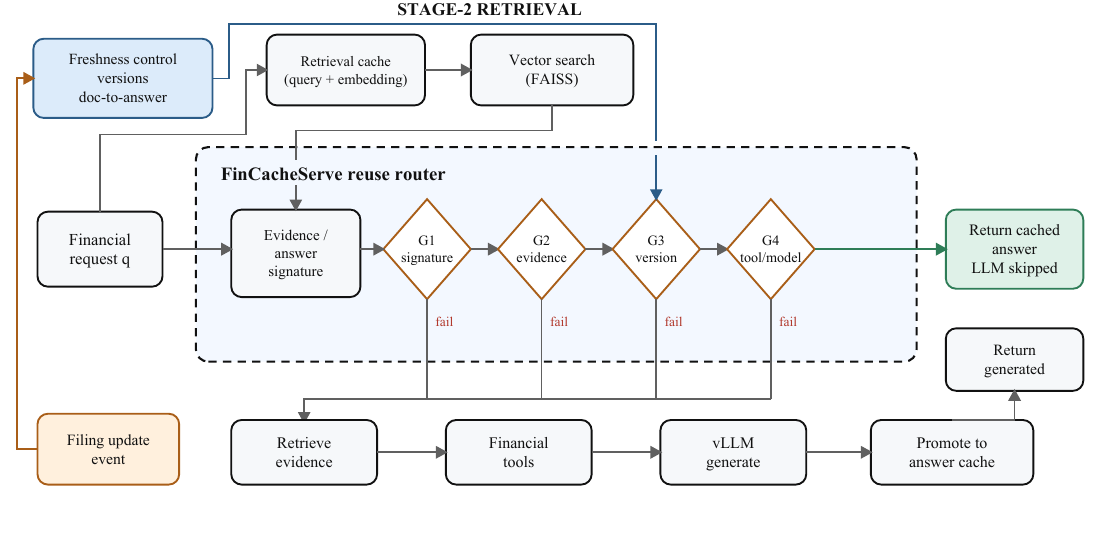}
  \caption{\sys serving architecture. A request reaches the answer-reuse router after retrieval-cache and vector-search resolution. The router admits reuse through financial signature, evidence, version, and tool/model gates. Filing updates update the version store and invalidate dependent answer entries before later cache reads.}
  \label{fig:architecture}
\end{figure*}

\subsection{Request signature}

The request signature contains company identity, reporting period, query family, document scope, tool requirement, model identity, and decoding configuration. Query family captures analytical intent, including liquidity, revenue change, debt, margins, risk-factor change, and cash-flow movement. Period identity prevents cross-quarter or cross-year reuse. Document scope separates annual reports, quarterly reports, current reports, and extracted XBRL facts. Tool requirement separates generated textual explanations from requests that depend on computed financial quantities.

\subsection{Answer-cache entry}

Each answer-cache entry stores answer text, cited document identifiers, chunk identifiers, chunk hashes, document versions, evidence fingerprint, tool fingerprint, model identity, decoding configuration, and request signature fields. The evidence fingerprint is an order-stable digest of cited chunk identifiers, chunk hashes, and document versions. The tool fingerprint hashes the tool name, tool version, normalized inputs, and returned structured payload. These hashes turn reusable answers into dependency-bound serving objects.

\subsection{Reuse gate}

The reuse gate first looks up exact and signature-compatible candidates. Candidate entries pass through four gate families. The signature gate checks company, query family, period, document scope, tool requirement, model identity, and decoding configuration. The evidence gate compares retrieved-support fingerprints. The version gate compares stored source-document versions against the current version store. The tool/model gate checks tool fingerprints and model configuration. A rejected candidate routes to retrieval, tool execution, model generation, and promotion into the answer cache.

\subsection{Dependency invalidation}

The metadata plane maintains a document-to-answer reverse index. When a filing update changes a document version, the system updates the version store and uses the reverse index to invalidate dependent answer entries. The lookup path is $O(1)$ for exact or bucketed signature candidates plus gate evaluation over a small candidate set. Invalidation is $O(k)$ for $k$ answer entries that depend on the updated document. The memory footprint is $O(N+E)$ for $N$ answer entries and $E$ document-to-answer edges.

\subsection{Admission and eviction under bounded capacity}

Hosted deployments bound answer-cache capacity by memory, privacy retention, and operational isolation. \sys therefore separates the correctness gate from the cache-management policy. The gate decides whether a candidate answer is dependency-consistent. The admission and eviction policy decides which generated answers deserve scarce cache space.

For an answer entry $c$, the default utility score is
\begin{equation}
U(c)=\hat{H}(s(c))\,C(c)-\lambda_m M(c)-\lambda_f F(c)-\lambda_u R(c),
\end{equation}
where $\hat{H}(s(c))$ estimates reuse demand for the normalized signature, $C(c)$ estimates avoided generation cost, $M(c)$ is entry size, $F(c)$ is document fan-out, and $R(c)$ is update hazard for the dependent documents. Entries with positive utility are admitted. When capacity is exhausted, \sys evicts the entry with the lowest utility density $U(c)/M(c)$. The policy runs in the metadata plane: it uses routing fields, document dependencies, and historical request counts already collected by the serving layer, and it leaves the dependency-consistency gate unchanged.

The offline CacheOpt policy estimates $\hat{H}$, $F$, and $R$ from the replayed trace to establish an upper-bound policy target short of Belady. The online CacheOpt policy estimates these terms from a sliding policy-history window of recent requests and filing updates. This separation makes the cache-management result auditable: global trace statistics define a policy ceiling, while the online variant measures deployable adaptation from serving history.

\subsection{Relationship to baseline cache policies}

Table~\ref{tab:novelty} positions \sys against cache families used in the evaluation. The central mechanism is the reuse contract: a cached answer remains eligible while its financial identity, evidence support, tool state, model state, and source versions remain compatible with the incoming request.

\begin{table*}[t]
\caption{Mechanism boundary relative to cache families evaluated in this paper.}
\label{tab:novelty}
\centering
\small
\begin{tabularx}{\textwidth}{@{}lYY@{}}
\toprule
Family & Reuse condition & \sys extension \\
\midrule
KV/prefix/context cache & Shared tokens or context states inside the model-serving path & Answer-level hit removes the hosted model call from the request path \\
TTL answer cache & Entry age within a fixed expiration window & Freshness tied to source-document versions and recorded dependencies \\
Semantic answer cache & Query similarity above a threshold & Similarity combined with financial identity, period, evidence, tool, model, and version checks \\
Grounded-style router & Evidence overlap and source support & Financial signature buckets plus reverse-index invalidation over mutable filing dependencies \\
Exact answer invalidation & Text-identical answer reuse with source-document invalidation & Paraphrase-level reuse for equivalent financial intent under the same dependency contract \\
\botrule
\end{tabularx}
\end{table*}

\section{Implementation}\label{sec:implementation}

The implementation contains five components: a request normalizer, retrieval cache, answer cache, metadata plane, and hosted-generation connector. Retrieval uses cached embeddings and FAISS search \cite{johnson2021faiss}. Generation uses vLLM-hosted Qwen2.5-Instruct models \cite{kwon2023vllm,yang2024qwen25}. The cache path records request-level routing, LLM invocation flags, skipped-call counters, token estimates, evidence hashes, tool hashes, version checks, stale-origin categories, and provider-call closure.

The answer cache maintains an exact index for repeated requests, a signature bucket index for financial-intent reuse, and a document-to-answer reverse index for invalidation. Gate decisions are logged per request. A cache hit records whether it is exact, signature, or semantic within a signature bucket. A cache miss records the rejected candidate family and the fallback generation path. Filing updates enter the metadata plane as version-store updates, followed by invalidation of all reverse-indexed answer entries.

The concurrency implementation uses a service-level metadata lock around version-store updates and answer-entry mutation. Hosted LLM calls execute outside the metadata critical section. The design preserves a single linearization order for cache reads and filing updates in the evaluated implementation.

\section{Experimental methodology}\label{sec:methodology}

\subsection{Workloads}

The evaluation uses SEC-derived financial-document workloads. Public filing content and metadata provide company, period, document scope, and source-version structure. Requests cover liquidity, revenue, margin, debt, risk, cash-flow, filing-change, and ratio-analysis families. Table~\ref{tab:workloads} summarizes the workloads.

\begin{table*}[t]
\caption{Evaluation workloads. Hosted workloads invoke vLLM; metadata stress runs the cache metadata path with deterministic fixtures.}
\label{tab:workloads}
\centering
\footnotesize
\renewcommand{\arraystretch}{0.92}
\setlength{\tabcolsep}{2.5pt}
\begin{tabularx}{\textwidth}{@{}>{\raggedright\arraybackslash}p{0.17\textwidth}>{\raggedright\arraybackslash}p{0.20\textwidth}>{\raggedright\arraybackslash}p{0.12\textwidth}X@{}}
\toprule
Workload & Requests & Model path & Purpose \\
\midrule
Main hosted trace & 2,230 & 7B & Full-trace provider-call accounting \\
Update stress & 208 & 7B & Paraphrases plus freshness events \\
Second trace mix & 172 & 7B & Seed and mix robustness \\
Model-size probes & 112 & 7B/14B/32B & Route behavior across model sizes \\
Operator suite & 544 & 32B & Strong-baseline comparison across three hosted seeds \\
Capacity replay & 600 warmup; 895 queries; 110 updates & det. & Bounded admission, eviction, oracle gap, update-rate, and policy-history sensitivity \\
Near-collision probes & 2,376 probes & det. & Cache-integrity stress across semantic, versioned, grounded, and dependency-consistent reuse \\
SLO and energy replay & 24 hosted metric files & det. & Dependency-fresh goodput, GPU-second accounting, and Wh sensitivity \\
Interleaved metadata stress & 4,096 queries; 512 updates/repeat & det. & Reverse-index invalidation, fan-out, and concurrency \\
Backend validation & 10k/100k entries; fan-out 1/16/128 & det. & In-memory and SQLite/WAL lookup, storage, and transactional invalidation \\
Normalizer audit & 500 annotated rows & offline & Human audit of reuse-signature metadata \\
\botrule
\end{tabularx}
\end{table*}

\subsection{Hosted execution environment}

Hosted generation experiments ran on AutoDL cloud GPU instances through a vLLM HTTP serving endpoint. The hosted 7B full-trace, update-stress, filing-QA, and concurrency runs used a single NVIDIA GeForce RTX 5090 GPU with 32,607 MiB of reported memory (32 GB class) serving Qwen2.5-7B-Instruct. The focused 14B model-size probes used the same RTX 5090-class hosted environment. The hosted 32B operator-suite runs used a single NVIDIA RTX PRO 6000 Blackwell Server Edition GPU with 97,887 MiB of reported memory (96 GB class) and NVIDIA driver 595.58.03, serving Qwen2.5-32B-Instruct. The deterministic capacity replay, near-collision probes, interleaved metadata stress, SQLite/WAL backend validation, SLO replay, energy replay, and normalizer audit were computed from archived traces or metadata fixtures and did not invoke a GPU. GPU-second and Wh results use archived per-request hosted timing metrics and explicit board-power assumptions; they are resource-sensitivity estimates rather than direct whole-system energy-meter readings.

\subsection{Baselines}

The evaluation compares \sys with no-cache serving, retrieval-cache serving, TTL answer caching, semantic answer caching without freshness checks, exact answer reuse with document invalidation, versioned semantic caching, entity-period semantic caching, grounded-style reuse, financial-signature reuse without freshness invalidation, and ablated \sys variants. The strongest safe semantic baseline combines semantic similarity with document-version checks. The grounded-style baseline checks evidence overlap and source support before reuse. The exact-invalidation baseline measures repeated-text reuse under document invalidation.

\subsection{Metrics}

The main efficiency metrics are LLM invocations, skipped LLM calls, skip rate, provider-call closure, prompt-token savings, completion-token savings, TTFT, end-to-end latency, throughput, SLO goodput, GPU seconds, and power-sensitivity estimates where available. SLO goodput counts requests that are error-free, dependency-fresh, and complete within a latency budget. Energy sensitivity reports Wh per 1,000 dependency-fresh SLO successes under explicit board-power assumptions. The cache-management replay additionally reports saved relative generation cost, oracle skip gap, eviction count, invalidation count, and lookup/invalidation p95 latency. The metadata-backend validation reports lookup p95, invalidation p95, storage bytes per entry, invalidated entries per update, and stale serves. The main consistency metrics are answer-cache stale serves, retrieval-evidence stale serves, tool-output stale serves, version mismatches, evidence-hash mismatches, and tool-hash mismatches. Wilson intervals are reported for selected proportion estimates where intervals are shown. Paired request-level comparisons report skip-rate differences when paired route-level outcomes are available.

\section{Results}\label{sec:results}

\subsection{Full hosted 7B trace}

Table~\ref{tab:main} reports the main hosted 7B trace. \sys skips 1,188 of 2,230 LLM calls (53.27\%) with zero observed dependency-stale outputs. The no-cache and retrieval-cache baselines invoke the model for every measured request. TTL and semantic no-freshness variants reach similar reuse levels but expose stale outputs after dependency updates.

\begin{table*}[t]
\caption{Full hosted Qwen2.5-7B trace. Dependency-stale counts aggregate answer-cache stale, retrieval-evidence stale, and tool-output stale categories.}
\label{tab:main}
\centering
\footnotesize
\setlength{\tabcolsep}{3pt}
\begin{tabular}{@{}lrrrr@{}}
\toprule
Variant & Requests & LLM calls & LLM skipped & Dep.-stale \\
\midrule
No-cache serving & 2,230 & 2,230 & 0 & 0 \\
Retrieval-cache serving & 2,230 & 2,230 & 0 & 1 \\
TTL answer cache & 2,230 & 1,057 & 1,173 & 1 \\
Semantic cache, no freshness & 2,230 & 1,057 & 1,173 & 1 \\
\sys & 2,230 & 1,042 & 1,188 & 0 \\
\botrule
\end{tabular}
\end{table*}

Figure~\ref{fig:efficiency} summarizes cost reduction and dependency-freshness behavior. The main trace establishes that answer-level reuse can remove more than half of hosted LLM calls under full provider-call accounting. The single stale output in retrieval-cache serving also motivates stale-origin reporting: stale dependencies can enter through cached retrieval evidence as well as answer-cache hits.

\begin{center}
  \centering
  \includegraphics[width=\columnwidth]{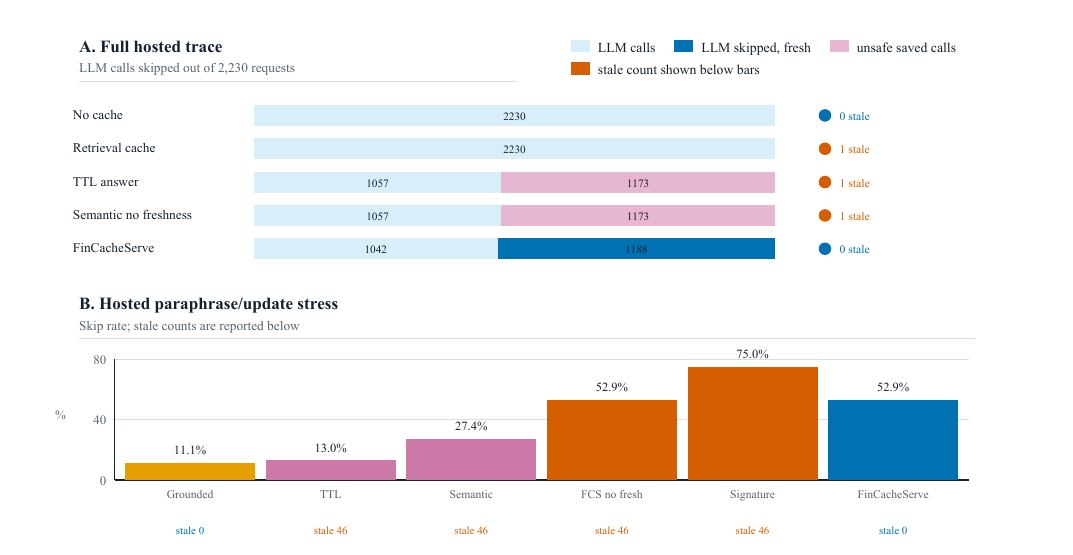}
  \captionof{figure}{LLM-call reduction and dependency-stale behavior across primary hosted conditions.}
  \label{fig:efficiency}
\end{center}

\subsection{Paraphrase and update stress}

The 208-request hosted stress workload exercises repeated financial intent, paraphrases, and filing-update events. \sys skips 52.88\% of LLM calls with zero dependency-stale outputs. Unsafe answer-cache controls expose the freshness cost of reuse without dependency invalidation: TTL, semantic no-freshness, signature no-freshness, and \sys without freshness invalidation each serve 46 dependency-stale outputs. Grounded-style safe reuse avoids stale outputs but skips 11.06\% of calls. The result separates two axes: aggressive answer reuse improves GPU cost, and dependency checks govern freshness under updates.

\begin{table*}[t]
\caption{Hosted paraphrase/update stress with Qwen2.5-7B.}
\label{tab:stress}
\centering
\scriptsize
\begin{tabular}{@{}lrrrr@{}}
\toprule
Variant & Requests & LLM skipped & Skip rate & Dep.-stale \\
\midrule
TTL answer cache & 208 & 27 & 12.98\% & 46 \\
Semantic cache, no freshness & 208 & 57 & 27.40\% & 46 \\
Signature cache, no freshness & 208 & 156 & 75.00\% & 46 \\
Grounded-style reuse & 208 & 23 & 11.06\% & 0 \\
\sys, no signature gate & 208 & 23 & 11.06\% & 0 \\
\sys, no freshness invalidation & 208 & 110 & 52.88\% & 46 \\
\sys & 208 & 110 & 52.88\% & 0 \\
\botrule
\end{tabular}
\end{table*}

\subsection{Strong 32B baselines across three hosted seeds}

Table~\ref{tab:operator} reports three hosted Qwen2.5-32B operator-suite seeds. \sys skips 290 of 544 requests (53.31\%, 95\% CI [49.11, 57.46]) with zero observed dependency-stale outputs. The strongest safe semantic baseline skips 212 requests (38.97\%), grounded-style reuse skips 122 (22.43\%), entity-period semantic caching skips 57 (10.48\%), and exact answer reuse with document invalidation skips 14 (2.57\%). Unsafe semantic or signature caches skip many calls but serve 126--134 dependency-stale outputs.

\begin{table*}[t]
\caption{Hosted 32B operator-suite comparison across three seeds. Paired $\Delta$ skip is \sys minus baseline in percentage points.}
\label{tab:operator}
\centering
\scriptsize
\setlength{\tabcolsep}{2pt}
\resizebox{\textwidth}{!}{%
\begin{tabular}{@{}lrrrrr@{}}
\toprule
Variant & Runs & Requests & LLM skip & Dep.-stale & Paired $\Delta$ skip \\
\midrule
\sys & 3 & 544 & 53.31\% [49.11, 57.46] & 0 & -- \\
Versioned semantic & 3 & 544 & 38.97\% [34.96, 43.13] & 0 & 14.34 \\
Grounded-style & 3 & 544 & 22.43\% [19.12, 26.12] & 0 & 30.88 \\
Entity-period semantic & 3 & 544 & 10.48\% [8.18, 13.33] & 0 & 42.83 \\
Exact + doc invalidation & 3 & 544 & 2.57\% [1.54, 4.27] & 0 & 50.74 \\
Semantic, no freshness & 3 & 544 & 50.55\% [46.36, 54.73] & 134 & 2.76 \\
Signature, no freshness & 3 & 544 & 76.47\% [72.73, 79.84] & 126 & -23.16 \\
TTL answer cache & 3 & 544 & 9.19\% [7.04, 11.91] & 126 & 44.12 \\
\sys, no signature gate & 3 & 544 & 10.48\% [8.18, 13.33] & 0 & 42.83 \\
\sys, no freshness invalidation & 3 & 544 & 53.31\% [49.11, 57.46] & 126 & 0.00 \\
\botrule
\end{tabular}
}
\end{table*}

The 32B result identifies the main performance mechanism. Removing the financial-signature layer reduces safe reuse to 10.48\%, matching the entity-period semantic baseline. Removing freshness invalidation preserves skip rate but exposes 126 dependency-stale outputs. The safe reuse gain therefore comes from the combination of financial-intent indexing and dependency invalidation.

\subsection{Capacity-limited admission and eviction}

Hosted answer caches operate under bounded memory. The 32B operator-suite traces therefore support a deterministic metadata-plane replay that preloads the same 120 warmup requests used by the hosted runs, replays 895 measured requests and 110 filing-update events across five archived traces, and varies answer-cache capacity. This replay evaluates cache-management decisions while keeping the dependency-consistency gate fixed.

\begin{figure}[!b]
  \centering
  \includegraphics[width=\columnwidth]{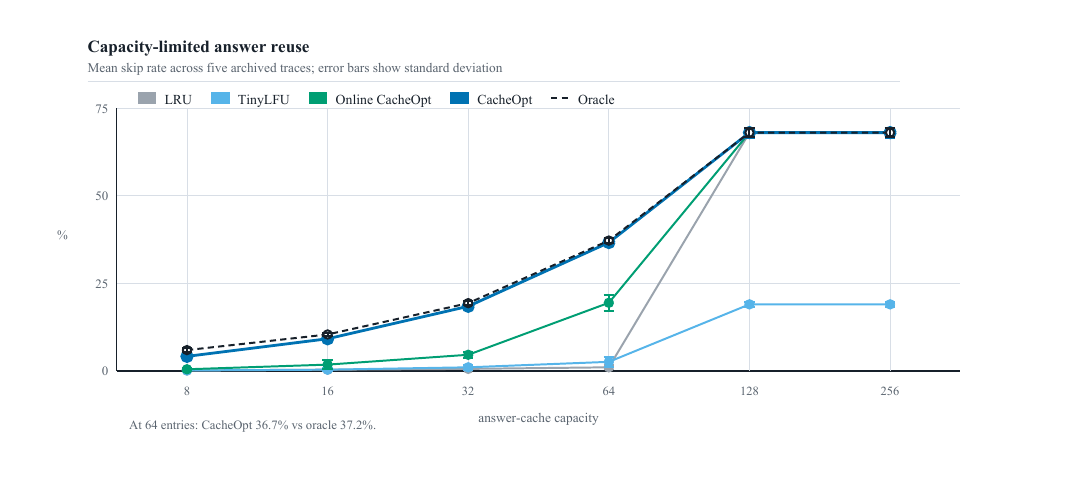}
  \caption{Capacity-limited metadata-plane replay. Utility-aware admission and eviction preserves reusable answer entries under bounded capacity and approaches the offline oracle at 32 and 64 entries. Error bars show standard deviation across five archived traces.}
  \label{fig:cacheopt_capacity}
\end{figure}

Figure~\ref{fig:cacheopt_capacity} shows that eviction policy dominates performance when the answer cache is smaller than the warmup working set. At 64 entries, LRU skips 1.01\% of measured calls and LFU skips 1.45\%; both policies evict entries that become useful during the measured phase. Utility-aware admission and eviction skips 36.66\% of calls with zero dependency-stale serves, within 0.56 percentage points of the Belady-style offline oracle. Online CacheOpt uses sliding-window request and update history. With the conservative 128-event window used in the main replay, it skips 19.44\% at the same capacity, exceeding LRU, LFU, and TinyLFU while retaining a 17.77 percentage-point oracle gap. At 128 entries, LRU, LFU, CacheOpt, Online CacheOpt, and the offline oracle all reach 68.06\% skipped calls; TinyLFU remains lower in this replay because its frequency sketch retains fewer measured-phase reusable entries.

\begin{center}
\begin{minipage}{\columnwidth}
\captionof{table}{Capacity-limited answer-cache replay over five archived 32B traces. Oracle gap is the skip-rate difference from the offline Belady-style oracle at the same capacity.}
\label{tab:cacheopt}
\centering
\footnotesize
\setlength{\tabcolsep}{4pt}
\resizebox{\columnwidth}{!}{%
\begin{tabular}{lrrrr}
\toprule
Policy & Capacity & Skip rate & Saved cost & Oracle gap \\
\midrule
FCS-LRU & 32 & 0.56\% & 0.56\% & 18.78\% \\
FCS-LRU & 64 & 1.01\% & 1.02\% & 36.21\% \\
FCS-LRU & 128 & 68.06\% & 68.00\% & 0.00\% \\
FCS-LFU & 32 & 0.67\% & 0.68\% & 18.67\% \\
FCS-LFU & 64 & 1.45\% & 1.47\% & 35.76\% \\
FCS-LFU & 128 & 68.06\% & 68.00\% & 0.00\% \\
FCS-TinyLFU & 32 & 1.01\% & 1.02\% & 18.33\% \\
FCS-TinyLFU & 64 & 2.57\% & 2.72\% & 34.65\% \\
FCS-TinyLFU & 128 & 19.00\% & 19.11\% & 49.07\% \\
FCS-CacheOpt & 32 & 18.44\% & 21.34\% & 0.90\% \\
FCS-CacheOpt & 64 & 36.66\% & 38.04\% & 0.56\% \\
FCS-CacheOpt & 128 & 68.06\% & 68.00\% & 0.00\% \\
FCS-OnlineCacheOpt & 32 & 4.59\% & 5.62\% & 14.74\% \\
FCS-OnlineCacheOpt & 64 & 19.44\% & 21.36\% & 17.77\% \\
FCS-OnlineCacheOpt & 128 & 68.06\% & 68.00\% & 0.00\% \\
Oracle-Belady & 32 & 19.33\% & 19.08\% & 0.00\% \\
Oracle-Belady & 64 & 37.22\% & 37.08\% & 0.00\% \\
Oracle-Belady & 128 & 68.06\% & 68.00\% & 0.00\% \\
\botrule
\end{tabular}
}
\end{minipage}
\end{center}

Figure~\ref{fig:cacheopt_online} reports the sliding-window online policy over replay windows. The first measured window contains many warmed entries and therefore has high reuse opportunity. Later windows contain fewer repeated signatures after invalidation, reducing the online hit rate. The result gives a deployable policy curve alongside the offline policy ceiling and separates deployable history-based adaptation from trace-wide utility estimation.

\begin{center}
  \begin{minipage}{\columnwidth}
    \centering
    \includegraphics[width=\columnwidth]{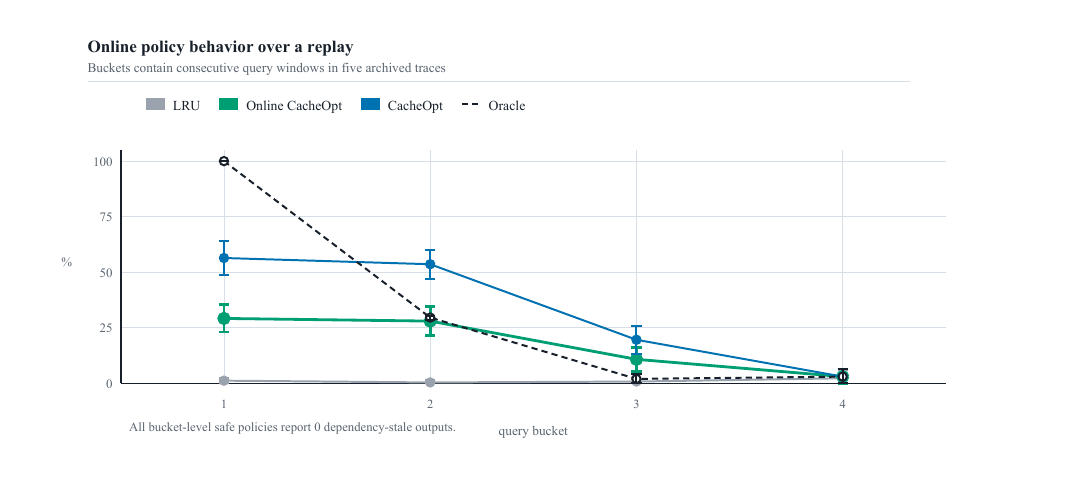}
    \captionof{figure}{Online CacheOpt behavior at 64 answer entries. The policy estimates reuse demand and update hazard from recent serving history. Error bars show standard deviation across five archived traces.}
    \label{fig:cacheopt_online}
  \end{minipage}
\end{center}

Figure~\ref{fig:cacheopt_window} varies the policy-history window at 64 entries. A 32-event window skips 36.10\% of calls with zero dependency-stale serves and a 1.12 percentage-point oracle gap. Windows of 128--512 events skip 19.44\%, reflecting older update history that suppresses admission for entries with near-term reuse. The scan preserves the distinction between the trace-wide offline policy, the deployable online policy, and the Belady-style oracle.

\begin{center}
  \begin{minipage}{\columnwidth}
    \centering
    \includegraphics[width=\columnwidth]{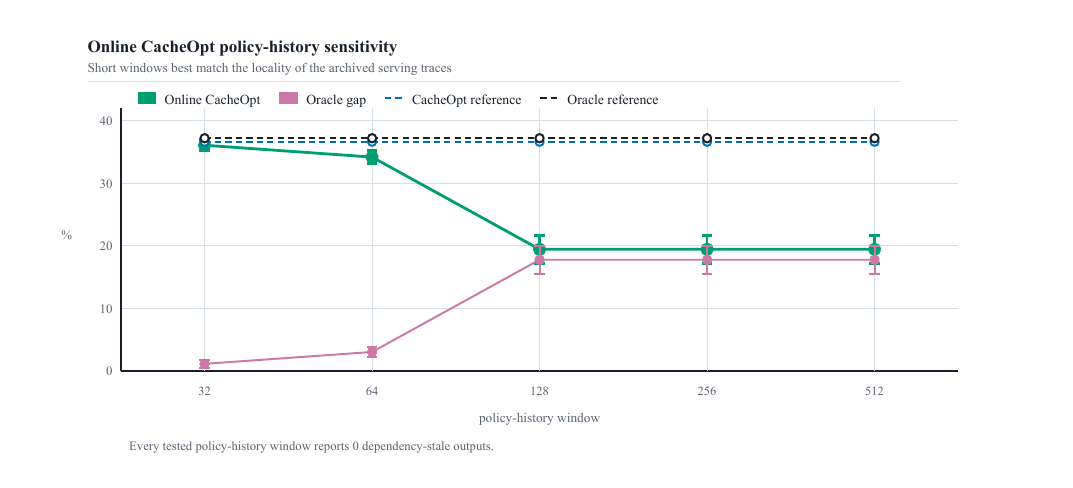}
    \captionof{figure}{Online CacheOpt policy-history sensitivity at 64 answer entries. The online policy remains dependency-fresh across all tested windows; the shortest tested window best matches the locality in the archived traces.}
    \label{fig:cacheopt_window}
  \end{minipage}
\end{center}

Figure~\ref{fig:cacheopt_update} varies the filing-update rate at 64 entries. Utility-aware cache management retains a stable advantage over LRU when update pressure is low or moderate, while dependency invalidation drives all safe policies toward lower reuse under high update pressure. The unsafe control skips 37.87\% of calls across update rates while serving stale dependencies in 22.78--32.17\% of measured requests, showing that invalidation is required for deployable call saving.

\begin{center}
  \begin{minipage}{\columnwidth}
    \centering
    \includegraphics[width=\columnwidth]{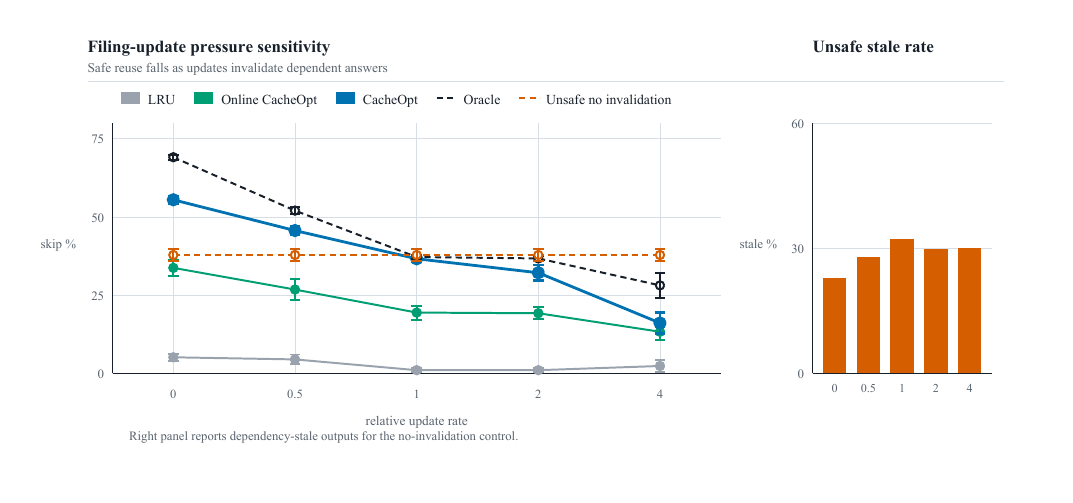}
    \captionof{figure}{Update-rate sensitivity at 64 answer entries. Higher filing-update pressure reduces safe answer reuse through dependency invalidation; the unsafe control preserves apparent reuse by serving stale dependencies.}
    \label{fig:cacheopt_update}
  \end{minipage}
\end{center}

\FloatBarrier

\subsection{Model-size sensitivity}

The matched model-size probes use Qwen2.5-7B, 14B, and 32B under the same route-level workloads. Workload structure and reuse policy determine skip behavior, while the monetary and latency value of each skipped call grows with model size. The 14B and 32B self-hosted vLLM runs also expose memory pressure and lower single-sequence serving throughput, strengthening the performance motivation for answer-level skipping on larger models.

\begin{center}
\begin{minipage}{\columnwidth}
\captionof{table}{Matched hosted model-size probes. Values summarize measured route-level requests.}
\label{tab:modelsize}
\centering
\footnotesize
\resizebox{\columnwidth}{!}{%
\begin{tabular}{@{}lrrrr@{}}
\toprule
Model & Requests & LLM skipped & Skip rate & Dep.-stale \\
\midrule
Qwen2.5-7B & 112 & 58 & 51.79\% & 0 \\
Qwen2.5-14B & 112 & 58 & 51.79\% & 0 \\
Qwen2.5-32B & 112 & 58 & 51.79\% & 0 \\
\botrule
\end{tabular}
}
\end{minipage}
\end{center}

\subsection{Hosted concurrency and throughput}

Figure~\ref{fig:concurrency} reports the hosted concurrency sweep over a bounded 53-request slice. \sys improves throughput over retrieval-cache serving by 1.58$\times$ on average across the tested concurrency settings while preserving zero observed dependency-stale outputs. The result indicates that skipped model calls translate into serving-path throughput under controlled concurrent query phases.

\begin{center}
  \centering
  \includegraphics[width=\columnwidth]{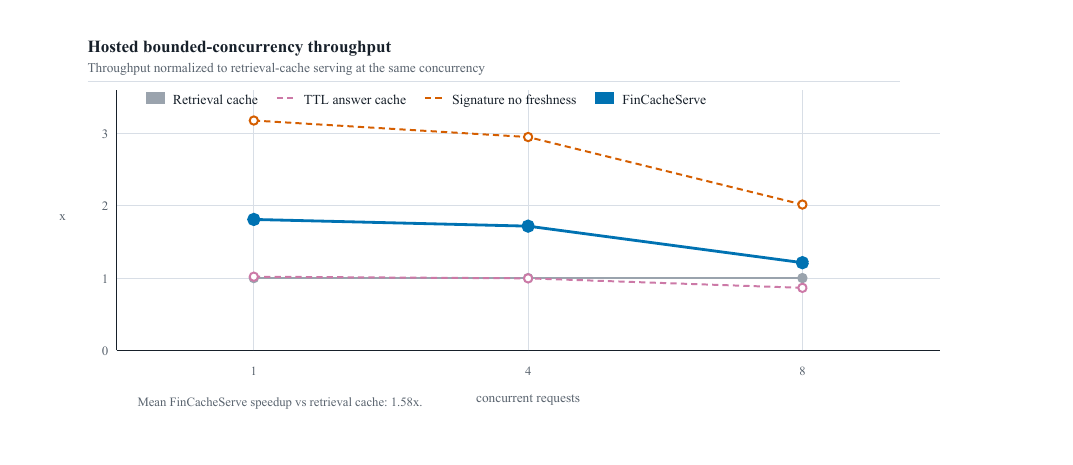}
  \captionof{figure}{Hosted throughput under bounded concurrent query phases. Throughput is normalized against retrieval-cache serving for the same phase.}
  \label{fig:concurrency}
\end{center}

\subsection{SLO-constrained goodput and GPU-second accounting}

The hosted 32B metrics also support a service-level replay over latency SLOs. Goodput counts requests that finish within the SLO, return without error, and serve dependency-fresh output. Figure~\ref{fig:slo_goodput} reports goodput at 0.25s--5s budgets. \sys reaches 53.31\% dependency-fresh goodput at every SLO up to 2s because answer-cache hits finish within the tight budgets while generation-path requests require roughly 3s. At 3s and 5s, generation-path requests increasingly satisfy the budget, raising \sys goodput to 77.02\% and 100\%.

\begin{center}
  \centering
  \includegraphics[width=\columnwidth]{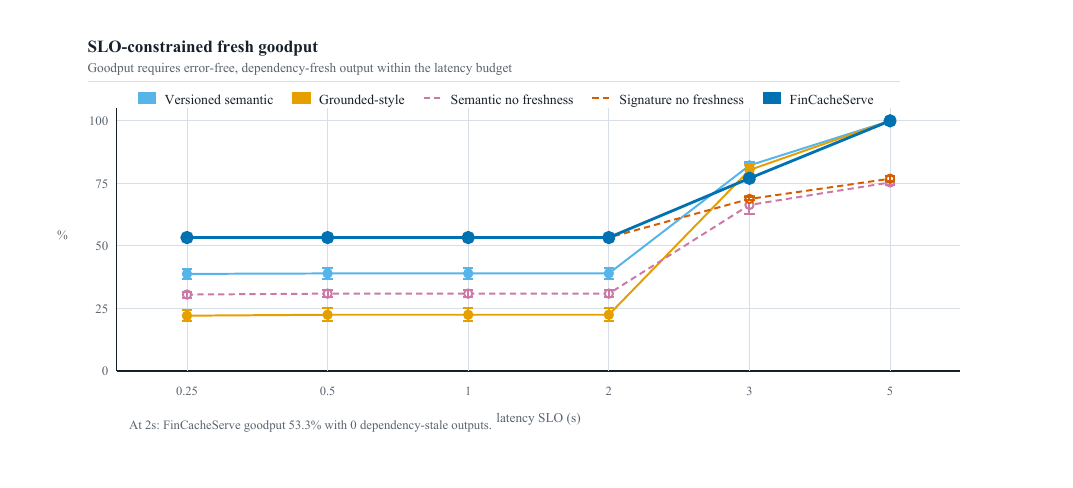}
  \captionof{figure}{SLO-constrained serving goodput from archived hosted 32B per-request metrics. Goodput counts error-free, dependency-fresh requests completed within the latency budget.}
  \label{fig:slo_goodput}
\end{center}

\begin{center}
\begin{minipage}{\columnwidth}
\captionof{table}{SLO replay at a 2s latency budget over hosted 32B metrics. GPU-sec/1k good is GPU seconds per 1,000 dependency-fresh SLO successes.}
\label{tab:slo}
\centering
\footnotesize
\resizebox{\columnwidth}{!}{%
\begin{tabular}{lrrrr}
\toprule
Variant & Goodput & Skip rate & Dep.-stale & GPU-sec/1k good \\
\midrule
Semantic, no freshness & 30.88\% & 50.54\% & 24.63\% & 40.45 \\
Versioned semantic & 38.96\% & 38.96\% & 0.00\% & 40.22 \\
Grounded-style reuse & 22.43\% & 22.43\% & 0.00\% & 89.22 \\
Signature, no freshness & 53.31\% & 76.48\% & 23.16\% & 11.27 \\
\sys & 53.31\% & 53.31\% & 0.00\% & 22.40 \\
\botrule
\end{tabular}
}
\end{minipage}
\end{center}

Table~\ref{tab:slo} shows the difference between apparent speed and deployable goodput. Signature no-freshness has the highest skip rate and lowest GPU seconds per SLO success, but 23.16\% of requests are dependency-stale. \sys matches its 2s goodput while eliminating observed dependency-stale output in the same hosted metrics.

Figure~\ref{fig:energy_cost} converts the same GPU-second estimates into Wh per 1,000 dependency-fresh 2s-SLO successes under explicit board-power assumptions. At 450 W, \sys uses 2.80 Wh per 1,000 fresh SLO successes, compared with 5.03 Wh for versioned semantic caching and 11.15 Wh for grounded-style reuse. This is a 44.30\% reduction relative to the strongest safe semantic baseline. The signature no-freshness control reaches 1.41 Wh but serves dependency-stale outputs in 23.16\% of requests, so its resource curve represents an unsafe negative control.

\begin{center}
\begin{minipage}{\columnwidth}
  \centering
  \includegraphics[width=\columnwidth]{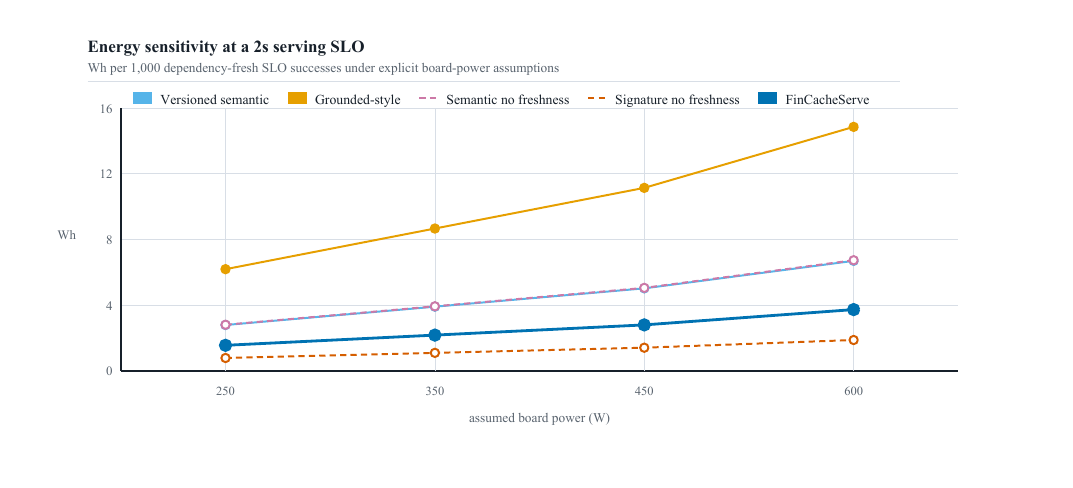}
  \captionof{figure}{Energy sensitivity from archived hosted 32B metrics at a 2s latency SLO. Wh values use measured GPU-second estimates and explicit board-power assumptions; unsafe baselines are shown as negative controls.}
  \label{fig:energy_cost}
\end{minipage}
\end{center}

\subsection{Field-level gate attribution}

Table~\ref{tab:field} reports controlled field-level gate ablations. The full gate admits the benign paraphrase and rejects every unsafe boundary case. Removing query family, period, tool hash, model identity, generation parameters, or evidence fingerprint exposes at least one unsafe reuse case. The result supports the inclusion of these fields in the reuse contract.

\begin{table*}[t]
\caption{Field-level gate ablation over controlled boundary probes.}
\label{tab:field}
\centering
\small
\begin{tabularx}{\textwidth}{@{}YrrY@{}}
\toprule
Variant & Safe accepts & Unsafe accepts & Exposed boundary \\
\midrule
Full gate & 1 & 0 & -- \\
Grounded-style, no financial signature & 1 & 2 & wrong period; wrong query family \\
Drop query family & 1 & 1 & wrong query family \\
Drop period key & 1 & 1 & wrong period \\
Drop tool hash & 1 & 1 & changed tool output \\
Drop model identity & 1 & 1 & changed model \\
Drop generation params & 1 & 1 & changed decoding parameters \\
Drop evidence fingerprint & 1 & 1 & changed evidence support \\
Drop company/scope/tool/source alone & 1 & 0 & overlapping gates reject the boundary probes \\
\botrule
\end{tabularx}
\end{table*}

\subsection{Controlled financial near-collision probes}

The field-level ablation isolates individual gates. A larger controlled probe suite evaluates full cache contracts under financial near-collisions. The suite contains 2,376 probes across issuers, periods, filing scopes, query families, tool-dependent families, evidence drifts, tool-output drifts, model changes, generation-parameter changes, and document-version updates. Each probe is semantically similar to a cached request; reuse is safe for benign paraphrases and unsafe for dependency-changing probes.

\begin{figure*}[t]
  \centering
  \includegraphics[width=0.88\textwidth]{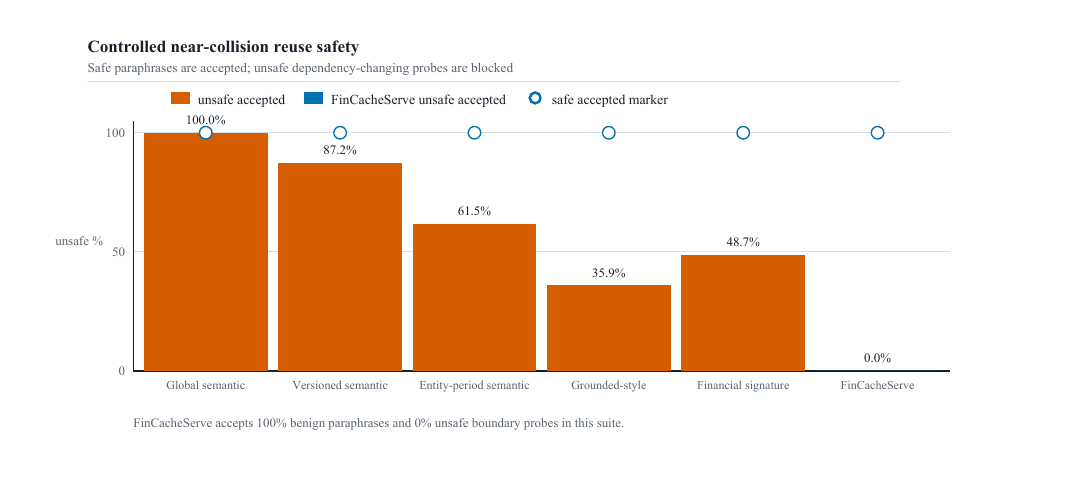}
  \caption{Unsafe accept rate on controlled financial near-collision probes. The suite isolates cache-contract boundaries and measures reuse-gate behavior under dependency-changing financial requests.}
  \label{fig:adversarial}
\end{figure*}

\begin{table*}[t]
\caption{Controlled near-collision cache-safety suite. Safe accept is the benign paraphrase accept rate; unsafe accept is the accept rate over dependency-changing probes.}
\label{tab:adversarial}
\centering
\small
\begin{tabular}{lrrrr}
\toprule
Policy & Cases & Safe accept & Unsafe accept & Unsafe block \\
\midrule
Global semantic & 2,376 & 100.00\% & 100.00\% & 0.00\% \\
Versioned semantic & 2,376 & 100.00\% & 87.18\% & 12.82\% \\
Entity-period semantic & 2,376 & 100.00\% & 61.54\% & 38.46\% \\
Grounded-style & 2,376 & 100.00\% & 35.90\% & 64.10\% \\
Financial signature & 2,376 & 100.00\% & 48.72\% & 51.28\% \\
\sys & 2,376 & 100.00\% & 0.00\% & 100.00\% \\
\botrule
\end{tabular}
\end{table*}

Table~\ref{tab:adversarial} explains why single-field cache contracts are insufficient for mutable financial RAG. Versioned semantic caching blocks version-update probes but accepts cross-company, period, scope, evidence, tool, model, and generation-parameter collisions. Grounded-style reuse blocks evidence and version collisions, yet it accepts model/configuration and several financial-identity drifts. \sys combines the financial signature, evidence/tool fingerprints, model/configuration identity, and version invalidation, blocking every unsafe probe while preserving benign paraphrase reuse in this controlled suite.

\subsection{Interleaved query/update consistency}

Table~\ref{tab:interleaved} reports the metadata-plane stress. Each setting runs five repeats with 4,096 queries and 512 filing updates per repeat. \sys records zero dependency-stale serves across 90 raw runs. The no-invalidation control records hundreds to more than one thousand stale serves per setting. Lookup and invalidation p95 latencies remain below 0.04 ms in the in-memory implementation.

\begin{center}
\begin{minipage}{\columnwidth}
\captionof{table}{Interleaved query/update consistency stress. Unsafe is the mean dependency-stale serves in the no-invalidation control.}
\label{tab:interleaved}
\centering
\scriptsize
\setlength{\tabcolsep}{2pt}
\resizebox{\columnwidth}{!}{%
\begin{tabular}{@{}rrrrrrr@{}}
\toprule
Entries & Hot set & QW & UW & \sys stale & Unsafe stale & p95 ms \\
\midrule
1k & 1k & 1  & 1 & 0.0 & 751.6  & 0.015 / 0.015 \\
1k & 1k & 16 & 2 & 0.0 & 767.0  & 0.017 / 0.010 \\
1k & 1k & 64 & 8 & 0.0 & 845.0  & 0.018 / 0.011 \\
10k & 1,024 & 1  & 1 & 0.0 & 1041.0 & 0.026 / 0.022 \\
10k & 1,024 & 16 & 2 & 0.0 & 1033.2 & 0.025 / 0.016 \\
10k & 1,024 & 64 & 8 & 0.0 & 922.8  & 0.026 / 0.018 \\
100k & 1,024 & 1  & 1 & 0.0 & 1081.6 & 0.025 / 0.025 \\
100k & 1,024 & 16 & 2 & 0.0 & 908.0  & 0.024 / 0.016 \\
100k & 1,024 & 64 & 8 & 0.0 & 905.2  & 0.030 / 0.020 \\
\botrule
\end{tabular}
}
\end{minipage}
\end{center}

\subsection{Transactional metadata-backend validation}

The interleaved stress above uses the in-memory metadata path from the hosted implementation. To check backend portability, a separate CPU validation maps the same answer-entry table, document-to-entry reverse index, and version store to SQLite/WAL. The workload covers 10k and 100k answer entries, fan-out 1, 16, and 128, three repeats per setting, 4,096 lookup operations, and up to 512 document-update transactions per run.

\begin{figure}[t]
  \centering
  \includegraphics[width=\columnwidth,height=0.38\textheight,keepaspectratio]{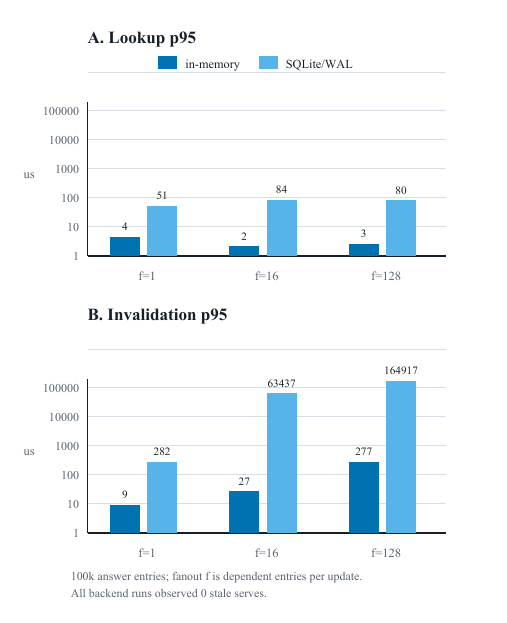}
  \caption{Metadata backend validation over in-memory and SQLite/WAL implementations.}
  \label{fig:metadata_backend}
\end{figure}

Figure~\ref{fig:metadata_backend} reports p95 lookup and invalidation latency for the in-memory and SQLite/WAL backends. At 100k entries and fan-out 16, in-memory lookup p95 is 2.13 $\mu$s and invalidation p95 is 27.00 $\mu$s. SQLite/WAL lookup p95 is 84.42 $\mu$s and transactional invalidation p95 is 63.44 ms while invalidating 16 entries per update on average. Across 36 raw backend runs, both backends record zero dependency-stale serves. The result confirms that the dependency contract can be implemented over a persistent transactional backend, while high fan-out updates introduce a measurable metadata-write cost.

\subsection{Normalizer and signature audit}

Table~\ref{tab:audit} reports the human audit of request metadata used by the reuse signature. The audit covers 500 annotated rows and excludes 13 rows without auditable query/request metadata, leaving 487 analyzable rows. Final signature-safe labels accept 384 rows and conservatively route 103 rows to generation. The audit records no hard contradiction in which a row is marked signature-safe while a component field is unsafe.

\begin{center}
\refstepcounter{table}\label{tab:audit}
\footnotesize
\noindent\begin{minipage}{\columnwidth}
\textbf{Table~\thetable} Human audit of metadata fields used for answer-reuse signatures.
\end{minipage}
\vspace{2pt}
\setlength{\tabcolsep}{3.0pt}
\begin{tabular}{@{}lrrr@{}}
\toprule
Field & Correct & Unsafe & Unresolved \\
\midrule
Company & 487 & 0 & 0 \\
Period & 452 & 27 & 8 \\
Query family & 487 & 0 & 0 \\
Document scope & 440 & 42 & 5 \\
Tool requirement & 423 & 64 & 0 \\
\midrule
Final signature safe & 384 & 103 & 0 \\
\botrule
\end{tabular}
\end{center}

The final signature-safe rate is 78.9\% with Wilson 95\% CI [75.0\%, 82.2\%]. Among analyzable rows, deterministic query-family consistency with the trace family is 87.7\% with Wilson 95\% CI [84.5\%, 90.3\%]. Rows marked unsafe by the audit become conservative generation routes, so metadata ambiguity primarily reduces reuse opportunity and preserves dependency freshness.

\section{Discussion}\label{sec:discussion}

\sys improves GPU-serving efficiency by moving reuse from retriever outputs or model-state reuse to dependency-checked answers. The 7B full trace establishes provider-call savings at full-run scale. The stress and operator-suite workloads show that the same mechanism keeps answer reuse fresh under filing updates. The 32B results strengthen the performance argument because the skipped call has higher serving value as model size increases. The capacity-limited replay adds a resource-management layer: a freshness-consistent cache also needs admission and eviction policies that preserve entries with future serving value when warmup traffic exceeds memory capacity. The SLO and energy replays translate skipped calls into dependency-fresh goodput and resource sensitivity, showing that low-latency success depends on both fast answers and dependency freshness. The SQLite/WAL validation adds a backend-realism check by mapping the metadata contract to persistent transactional storage.

The results also clarify the role of each mechanism. Financial-signature indexing recovers paraphrased analytical intent that exact matching misses. Versioned invalidation prevents stale reuse after filing updates. Evidence and tool fingerprints protect boundary cases where document versions alone leave ambiguity. Model identity and decoding configuration keep answer reuse aligned with the serving policy used at admission. The near-collision suite shows that these fields act as a joint contract: semantic similarity, version checks, evidence overlap, or financial signatures alone leave unsafe acceptance paths. The complete contract creates a compact reuse object that can be audited, logged, and invalidated.

The evaluation separates conservative reuse, unsafe reuse, and dependency-consistent answer reuse. Exact reuse and grounded-style safe reuse preserve dependency freshness but leave many repeated financial requests on the GPU path. Semantic and signature no-freshness variants save calls aggressively and expose stale outputs. \sys preserves most of the safe answer-level savings while converting update-sensitive cases into recomputation. The CacheOpt replay further shows that simple recency policies can discard the reusable portion of a warmed answer cache under tight capacity. Trace-wide utility estimates approach the offline oracle, and sliding-window Online CacheOpt provides a deployable policy curve with measurable oracle gap. The policy-history scan shows that online utility estimation is sensitive to workload locality; short recent-history windows recover near-offline reuse on the archived traces, while longer windows remain safe and more conservative.

\section{Validity scope}\label{sec:validity}

The evidence covers public SEC-derived workloads, hosted vLLM execution, Qwen2.5 models, an in-memory metadata plane with service-level linearization, and a SQLite/WAL backend validation for persistent metadata operations. The consistency claim is dependency freshness with respect to recorded source versions, evidence fingerprints, and tool fingerprints. Factual financial correctness remains a separate application-quality property and requires source review, answer audit, and domain governance in deployment.

The hosted traces include repeated analytical requests and generated paraphrase/update stress over real filing-derived dependencies. The near-collision probes are controlled mechanism tests that isolate dependency-changing boundary conditions. Production traffic can differ in query diversity, update cadence, access skew, tool composition, SLO distribution, and cache capacity. The metadata replay evaluates capacity, update-rate, and policy-history sensitivity over archived traces. The metadata stress validates the reverse-index and invalidation path up to 100k entries and 64 query workers in the evaluated implementation; the SQLite/WAL validation shows transactional backend portability and high fan-out invalidation cost. Energy estimates use GPU-second traces and explicit board-power assumptions. Distributed cache deployment, multi-replica invalidation, and cross-region consistency require additional engineering validation.

\section{Conclusion}\label{sec:conclusion}

\sys demonstrates dependency-consistent answer reuse for GPU-efficient RAG serving over mutable financial documents. The system treats generated answers as cacheable serving objects with explicit dependencies on financial identity, evidence support, tool outputs, model state, decoding configuration, and source-document versions. On a 2,230-request hosted 7B trace, \sys skips 53.27\% of LLM calls with zero observed dependency-stale outputs. Across three hosted 32B operator-suite seeds, it skips 53.31\% of requests, exceeding versioned semantic caching by 14.34 percentage points and grounded-style reuse by 30.88 percentage points under the same dependency-stale criterion. Under a 64-entry answer-cache budget, utility-aware admission skips 36.66\% of measured calls and stays within 0.56 percentage points of an offline oracle; a 32-event Online CacheOpt window skips 36.10\% with a 1.12 percentage-point oracle gap. A 2,376-probe near-collision suite blocks every unsafe candidate under the complete reuse contract, and a hosted-metrics replay reaches 53.31\% dependency-fresh goodput at a 2s SLO. At 450 W, the same hosted metrics imply 2.80 Wh per 1,000 dependency-fresh 2s-SLO successes, 44.30\% below versioned semantic caching. The metadata-plane experiments reach 100k-entry indexes, 64 query workers in interleaved stress, and SQLite/WAL transactional invalidation with zero observed dependency-stale serves. The results support answer-level caching as a practical performance mechanism for RAG services when reuse is coupled to dependency tracking, capacity-aware admission, and update-driven invalidation.

\raggedbottom
\backmatter

\section*{Supplementary information}
No supplementary artifact is submitted with this arXiv version.

\section*{Acknowledgements}
The authors thank collaborators who provided early feedback on the problem setting and figure design.

\section*{Declarations}

\subsection*{Funding}
No external funding was used for this study.

\subsection*{Conflict of interest}
The authors declare no competing interests.

\subsection*{Ethics approval and consent to participate}
Not applicable.

\subsection*{Data availability}
No data artifact is submitted with this arXiv version. The source filings are public SEC EDGAR documents. A curated reproducibility package with accession identifiers, derived traces, summaries, and checksums may be prepared for a future venue or archival release.

\subsection*{Code availability}
No code artifact is submitted with this arXiv version. A curated reproducibility package with workload reconstruction, cache-policy replay, trace validation, and table/figure generation code may be prepared for a future venue or archival release.

\subsection*{Author contribution}
Lingteng Zeng: conceptualization, methodology, software, validation, formal analysis, investigation, data curation, visualization, writing, and artifact preparation. Yifan Jin: data curation, validation, and investigation.

\balance
\bibliography{refs_arxiv}


\begin{thebibliography}{21}
\ifx \bisbn   \undefined \def \bisbn  #1{ISBN #1}\fi
\ifx \binits  \undefined \def \binits#1{#1}\fi
\ifx \bauthor  \undefined \def \bauthor#1{#1}\fi
\ifx \batitle  \undefined \def \batitle#1{#1}\fi
\ifx \bjtitle  \undefined \def \bjtitle#1{#1}\fi
\ifx \bvolume  \undefined \def \bvolume#1{\textbf{#1}}\fi
\ifx \byear  \undefined \def \byear#1{#1}\fi
\ifx \bissue  \undefined \def \bissue#1{#1}\fi
\ifx \bfpage  \undefined \def \bfpage#1{#1}\fi
\ifx \blpage  \undefined \def \blpage #1{#1}\fi
\ifx \burl  \undefined \def \burl#1{\textsf{#1}}\fi
\ifx \doiurl  \undefined \def \doiurl#1{\url{https://doi.org/#1}}\fi
\ifx \betal  \undefined \def \betal{\textit{et al.}}\fi
\ifx \binstitute  \undefined \def \binstitute#1{#1}\fi
\ifx \binstitutionaled  \undefined \def \binstitutionaled#1{#1}\fi
\ifx \bctitle  \undefined \def \bctitle#1{#1}\fi
\ifx \beditor  \undefined \def \beditor#1{#1}\fi
\ifx \bpublisher  \undefined \def \bpublisher#1{#1}\fi
\ifx \bbtitle  \undefined \def \bbtitle#1{#1}\fi
\ifx \bedition  \undefined \def \bedition#1{#1}\fi
\ifx \bseriesno  \undefined \def \bseriesno#1{#1}\fi
\ifx \blocation  \undefined \def \blocation#1{#1}\fi
\ifx \bsertitle  \undefined \def \bsertitle#1{#1}\fi
\ifx \bsnm \undefined \def \bsnm#1{#1}\fi
\ifx \bsuffix \undefined \def \bsuffix#1{#1}\fi
\ifx \bparticle \undefined \def \bparticle#1{#1}\fi
\ifx \barticle \undefined \def \barticle#1{#1}\fi
\bibcommenthead
\ifx \bconfdate \undefined \def \bconfdate #1{#1}\fi
\ifx \botherref \undefined \def \botherref #1{#1}\fi
\ifx \url \undefined \def \url#1{\textsf{#1}}\fi
\ifx \bchapter \undefined \def \bchapter#1{#1}\fi
\ifx \bbook \undefined \def \bbook#1{#1}\fi
\ifx \bcomment \undefined \def \bcomment#1{#1}\fi
\ifx \oauthor \undefined \def \oauthor#1{#1}\fi
\ifx \citeauthoryear \undefined \def \citeauthoryear#1{#1}\fi
\ifx \endbibitem  \undefined \def \endbibitem {}\fi
\ifx \bconflocation  \undefined \def \bconflocation#1{#1}\fi
\ifx \arxivurl  \undefined \def \arxivurl#1{\textsf{#1}}\fi
\csname PreBibitemsHook\endcsname

%%% 1
\bibitem[\protect\citeauthoryear{Kwon et~al.}{2023}]{kwon2023vllm}
\begin{bchapter}
\bauthor{\bsnm{Kwon}, \binits{W.}},
\bauthor{\bsnm{Li}, \binits{Z.}},
\bauthor{\bsnm{Zhuang}, \binits{S.}},
\bauthor{\bsnm{Sheng}, \binits{Y.}},
\bauthor{\bsnm{Zheng}, \binits{L.}},
\bauthor{\bsnm{Yu}, \binits{C.H.}},
\bauthor{\bsnm{Gonzalez}, \binits{J.E.}},
\bauthor{\bsnm{Zhang}, \binits{H.}},
\bauthor{\bsnm{Stoica}, \binits{I.}}:
\bctitle{Efficient memory management for large language model serving with
  {PagedAttention}}.
In: \bbtitle{Proceedings of the 29th ACM Symposium on Operating Systems
  Principles},
pp. \bfpage{611}--\blpage{626}
(\byear{2023}).
\doiurl{10.1145/3600006.3613165}
\end{bchapter}
\endbibitem

%%% 2
\bibitem[\protect\citeauthoryear{Zheng et~al.}{2023}]{zheng2023sglang}
\begin{botherref}
\oauthor{\bsnm{Zheng}, \binits{L.}},
\oauthor{\bsnm{Yin}, \binits{L.}},
\oauthor{\bsnm{Xie}, \binits{Z.}},
\oauthor{\bsnm{Sun}, \binits{C.}},
\oauthor{\bsnm{Huang}, \binits{J.}},
\oauthor{\bsnm{Yu}, \binits{C.H.}},
\oauthor{\bsnm{Cao}, \binits{S.}},
\oauthor{\bsnm{Kozyrakis}, \binits{C.}},
\oauthor{\bsnm{Stoica}, \binits{I.}},
\oauthor{\bsnm{Gonzalez}, \binits{J.E.}},
\oauthor{\bsnm{Barrett}, \binits{C.}},
\oauthor{\bsnm{Sheng}, \binits{Y.}}:
{SGLang}: Efficient Execution of Structured Language Model Programs
(2023).
\doiurl{10.48550/arXiv.2312.07104}
\end{botherref}
\endbibitem

%%% 3
\bibitem[\protect\citeauthoryear{Jin et~al.}{2024}]{jin2024ragcache}
\begin{botherref}
\oauthor{\bsnm{Jin}, \binits{C.}},
\oauthor{\bsnm{Zhang}, \binits{Z.}},
\oauthor{\bsnm{Jiang}, \binits{X.}},
\oauthor{\bsnm{Liu}, \binits{F.}},
\oauthor{\bsnm{Liu}, \binits{X.}},
\oauthor{\bsnm{Liu}, \binits{X.}},
\oauthor{\bsnm{Jin}, \binits{X.}}:
{RAGCache}: Efficient Knowledge Caching for Retrieval-Augmented Generation
(2024).
\doiurl{10.48550/arXiv.2404.12457}
\end{botherref}
\endbibitem

%%% 4
\bibitem[\protect\citeauthoryear{Yao et~al.}{2024}]{yao2024cacheblend}
\begin{botherref}
\oauthor{\bsnm{Yao}, \binits{J.}},
\oauthor{\bsnm{Li}, \binits{H.}},
\oauthor{\bsnm{Liu}, \binits{Y.}},
\oauthor{\bsnm{Ray}, \binits{S.}},
\oauthor{\bsnm{Cheng}, \binits{Y.}},
\oauthor{\bsnm{Zhang}, \binits{Q.}},
\oauthor{\bsnm{Du}, \binits{K.}},
\oauthor{\bsnm{Lu}, \binits{S.}},
\oauthor{\bsnm{Jiang}, \binits{J.}}:
{CacheBlend}: Fast Large Language Model Serving for {RAG} with Cached Knowledge
  Fusion
(2024).
\doiurl{10.48550/arXiv.2405.16444}
\end{botherref}
\endbibitem

%%% 5
\bibitem[\protect\citeauthoryear{Jiang et~al.}{2025}]{jiang2025contextpilot}
\begin{botherref}
\oauthor{\bsnm{Jiang}, \binits{Y.}},
\oauthor{\bsnm{Huang}, \binits{Y.}},
\oauthor{\bsnm{Cheng}, \binits{L.}},
\oauthor{\bsnm{Deng}, \binits{C.}},
\oauthor{\bsnm{Sun}, \binits{X.}},
\oauthor{\bsnm{Mai}, \binits{L.}}:
{ContextPilot}: Fast Long-Context Inference via Context Reuse
(2025).
\doiurl{10.48550/arXiv.2511.03475}
\end{botherref}
\endbibitem

%%% 6
\bibitem[\protect\citeauthoryear{Feng et~al.}{2025}]{feng2025adaptcache}
\begin{botherref}
\oauthor{\bsnm{Feng}, \binits{S.}},
\oauthor{\bsnm{Li}, \binits{H.}},
\oauthor{\bsnm{Du}, \binits{K.}},
\oauthor{\bsnm{Gu}, \binits{Z.}},
\oauthor{\bsnm{Liu}, \binits{Y.}},
\oauthor{\bsnm{Yao}, \binits{J.}},
\oauthor{\bsnm{Ray}, \binits{S.}},
\oauthor{\bsnm{Shen}, \binits{S.}},
\oauthor{\bsnm{Cheng}, \binits{Y.}},
\oauthor{\bsnm{Ananthanarayanan}, \binits{G.}},
\oauthor{\bsnm{Jiang}, \binits{J.}}:
{AdaptCache}: {KV} Cache Native Storage Hierarchy for Low-Delay and
  High-Quality Language Model Serving.
Accepted at SOSP 2025 BigMem Workshop
(2025).
\doiurl{10.48550/arXiv.2509.00105}
\end{botherref}
\endbibitem

%%% 7
\bibitem[\protect\citeauthoryear{Johnson et~al.}{2021}]{johnson2021faiss}
\begin{barticle}
\bauthor{\bsnm{Johnson}, \binits{J.}},
\bauthor{\bsnm{Douze}, \binits{M.}},
\bauthor{\bsnm{J{\'e}gou}, \binits{H.}}:
\batitle{Billion-scale similarity search with {GPU}s}.
\bjtitle{IEEE Transactions on Big Data}
\bvolume{7}(\bissue{3}),
\bfpage{535}--\blpage{547}
(\byear{2021})
\doiurl{10.1109/TBDATA.2019.2921572}
\end{barticle}
\endbibitem

%%% 8
\bibitem[\protect\citeauthoryear{Yang et~al.}{2024}]{yang2024qwen25}
\begin{botherref}
\oauthor{\bsnm{Yang}, \binits{A.}},
\oauthor{\bsnm{Yang}, \binits{B.}},
\oauthor{\bsnm{Zhang}, \binits{B.}},
\oauthor{\bsnm{Hui}, \binits{B.}},
\oauthor{\bsnm{Zheng}, \binits{B.}},
\oauthor{\bsnm{Yu}, \binits{B.}},
\oauthor{\bsnm{Li}, \binits{C.}},
\oauthor{\bsnm{Liu}, \binits{D.}},
\oauthor{\bsnm{Huang}, \binits{F.}},
\oauthor{\bsnm{Wei}, \binits{H.}},
\oauthor{\bsnm{Lin}, \binits{H.}}, et al.:
{Qwen2.5} Technical Report
(2024).
\doiurl{10.48550/arXiv.2412.15115}
\end{botherref}
\endbibitem

%%% 9
\bibitem[\protect\citeauthoryear{Chen et~al.}{2021}]{chen2021finqa}
\begin{bchapter}
\bauthor{\bsnm{Chen}, \binits{Z.}},
\bauthor{\bsnm{Chen}, \binits{W.}},
\bauthor{\bsnm{Smiley}, \binits{C.}},
\bauthor{\bsnm{Shah}, \binits{S.}},
\bauthor{\bsnm{Borova}, \binits{I.}},
\bauthor{\bsnm{Langdon}, \binits{D.}},
\bauthor{\bsnm{Moussa}, \binits{R.}},
\bauthor{\bsnm{Beane}, \binits{M.}},
\bauthor{\bsnm{Huang}, \binits{T.-H.}},
\bauthor{\bsnm{Routledge}, \binits{B.}},
\bauthor{\bsnm{Wang}, \binits{W.Y.}}:
\bctitle{{FinQA}: A dataset of numerical reasoning over financial data}.
In: \bbtitle{Proceedings of the 2021 Conference on Empirical Methods in Natural
  Language Processing},
pp. \bfpage{3697}--\blpage{3711}
(\byear{2021}).
\doiurl{10.18653/v1/2021.emnlp-main.300}
\end{bchapter}
\endbibitem

%%% 10
\bibitem[\protect\citeauthoryear{Islam et~al.}{2023}]{islam2023financebench}
\begin{botherref}
\oauthor{\bsnm{Islam}, \binits{P.}},
\oauthor{\bsnm{Kannappan}, \binits{A.}},
\oauthor{\bsnm{Kiela}, \binits{D.}},
\oauthor{\bsnm{Qian}, \binits{R.}},
\oauthor{\bsnm{Scherrer}, \binits{N.}},
\oauthor{\bsnm{Vidgen}, \binits{B.}}:
{FinanceBench}: A New Benchmark for Financial Question Answering
(2023).
\doiurl{10.48550/arXiv.2311.11944}
\end{botherref}
\endbibitem

%%% 11
\bibitem[\protect\citeauthoryear{{U.S. Securities and Exchange
  Commission}}{2025}]{sec2025edgar}
\begin{botherref}
\oauthor{\bsnm{{U.S. Securities and Exchange Commission}}}:
{EDGAR} Application Programming Interfaces ({APIs}).
\url{https://www.sec.gov/search-filings/edgar-application-programming-interfaces}.
Last reviewed or updated April 8, 2025; accessed June 29, 2026
(2025)
\end{botherref}
\endbibitem

%%% 12
\bibitem[\protect\citeauthoryear{Liu et~al.}{2025}]{liu2025semanticcache}
\begin{botherref}
\oauthor{\bsnm{Liu}, \binits{X.}},
\oauthor{\bsnm{Atalar}, \binits{B.}},
\oauthor{\bsnm{Dai}, \binits{X.}},
\oauthor{\bsnm{Zuo}, \binits{J.}},
\oauthor{\bsnm{Wang}, \binits{S.}},
\oauthor{\bsnm{Lui}, \binits{J.C.S.}},
\oauthor{\bsnm{Chen}, \binits{W.}},
\oauthor{\bsnm{Joe-Wong}, \binits{C.}}:
Semantic Caching for Low-Cost {LLM} Serving: From Offline Learning to Online
  Adaptation.
Accepted to IEEE INFOCOM 2026
(2025).
\doiurl{10.48550/arXiv.2508.07675}
\end{botherref}
\endbibitem

%%% 13
\bibitem[\protect\citeauthoryear{Singh et~al.}{2026}]{singh2026krites}
\begin{botherref}
\oauthor{\bsnm{Singh}, \binits{A.K.}},
\oauthor{\bsnm{Wang}, \binits{H.}},
\oauthor{\bsnm{Attaluri}, \binits{L.N.S.}},
\oauthor{\bsnm{Chiam}, \binits{T.}},
\oauthor{\bsnm{Zhu}, \binits{W.}}:
Asynchronous Verified Semantic Caching for Tiered {LLM} Architectures
(2026).
\doiurl{10.48550/arXiv.2602.13165}
\end{botherref}
\endbibitem

%%% 14
\bibitem[\protect\citeauthoryear{Shah}{2026}]{shah2026groundedcache}
\begin{botherref}
\oauthor{\bsnm{Shah}, \binits{S.H.}}:
Grounded Cache Routing for Retrieval-Augmented Generation: When Is It Safe to
  Reuse an Answer?
(2026).
\doiurl{10.48550/arXiv.2605.27494}
\end{botherref}
\endbibitem

%%% 15
\bibitem[\protect\citeauthoryear{Haqiq et~al.}{2025}]{haqiq2025mincache}
\begin{barticle}
\bauthor{\bsnm{Haqiq}, \binits{K.}},
\bauthor{\bsnm{Jahan}, \binits{M.V.}},
\bauthor{\bsnm{Farimani}, \binits{S.A.}},
\bauthor{\bsnm{Masoom}, \binits{S.M.F.}}:
\batitle{{MinCache}: A hybrid cache system for efficient chatbots with
  hierarchical embedding matching and {LLM}}.
\bjtitle{Future Generation Computer Systems}
\bvolume{170},
\bfpage{107822}
(\byear{2025})
\doiurl{10.1016/j.future.2025.107822}
\end{barticle}
\endbibitem

%%% 16
\bibitem[\protect\citeauthoryear{Nottingham
  et~al.}{2022}]{nottingham2022rfc9111}
\begin{botherref}
\oauthor{\bsnm{Nottingham}, \binits{M.}},
\oauthor{\bsnm{Fielding}, \binits{R.T.}},
\oauthor{\bsnm{Reschke}, \binits{J.}}:
{HTTP} Caching.
RFC 9111
(2022).
\doiurl{10.17487/RFC9111} .
\url{https://www.rfc-editor.org/rfc/rfc9111}
\end{botherref}
\endbibitem

%%% 17
\bibitem[\protect\citeauthoryear{Gupta and
  Mumick}{1995}]{gupta1995materialized}
\begin{barticle}
\bauthor{\bsnm{Gupta}, \binits{A.}},
\bauthor{\bsnm{Mumick}, \binits{I.S.}}:
\batitle{Maintenance of materialized views: Problems, techniques, and
  applications}.
\bjtitle{IEEE Data Engineering Bulletin}
\bvolume{18}(\bissue{2}),
\bfpage{3}--\blpage{18}
(\byear{1995})
\end{barticle}
\endbibitem

%%% 18
\bibitem[\protect\citeauthoryear{Mistry et~al.}{2001}]{mistry2001materialized}
\begin{bchapter}
\bauthor{\bsnm{Mistry}, \binits{H.}},
\bauthor{\bsnm{Roy}, \binits{P.}},
\bauthor{\bsnm{Sudarshan}, \binits{S.}},
\bauthor{\bsnm{Ramamritham}, \binits{K.}}:
\bctitle{Materialized view selection and maintenance using multi-query
  optimization}.
In: \bbtitle{Proceedings of the 2001 {ACM} {SIGMOD} International Conference on
  Management of Data},
pp. \bfpage{307}--\blpage{318}
(\byear{2001}).
\doiurl{10.1145/375663.375706}
\end{bchapter}
\endbibitem

%%% 19
\bibitem[\protect\citeauthoryear{Buneman et~al.}{2001}]{buneman2001whywhere}
\begin{bchapter}
\bauthor{\bsnm{Buneman}, \binits{P.}},
\bauthor{\bsnm{Khanna}, \binits{S.}},
\bauthor{\bsnm{Tan}, \binits{W.-C.}}:
\bctitle{Why and where: A characterization of data provenance}.
In: \bbtitle{Database Theory -- {ICDT} 2001},
pp. \bfpage{316}--\blpage{330}
(\byear{2001}).
\doiurl{10.1007/3-540-44503-X_20}
\end{bchapter}
\endbibitem

%%% 20
\bibitem[\protect\citeauthoryear{DeCandia et~al.}{2007}]{decandia2007dynamo}
\begin{bchapter}
\bauthor{\bsnm{DeCandia}, \binits{G.}},
\bauthor{\bsnm{Hastorun}, \binits{D.}},
\bauthor{\bsnm{Jampani}, \binits{M.}},
\bauthor{\bsnm{Kakulapati}, \binits{G.}},
\bauthor{\bsnm{Lakshman}, \binits{A.}},
\bauthor{\bsnm{Pilchin}, \binits{A.}},
\bauthor{\bsnm{Sivasubramanian}, \binits{S.}},
\bauthor{\bsnm{Vosshall}, \binits{P.}},
\bauthor{\bsnm{Vogels}, \binits{W.}}:
\bctitle{Dynamo: Amazon's highly available key-value store}.
In: \bbtitle{Proceedings of Twenty-First {ACM} {SIGOPS} Symposium on Operating
  Systems Principles},
pp. \bfpage{205}--\blpage{220}
(\byear{2007}).
\doiurl{10.1145/1294261.1294281}
\end{bchapter}
\endbibitem

%%% 21
\bibitem[\protect\citeauthoryear{Cooper et~al.}{2008}]{cooper2008pnuts}
\begin{bchapter}
\bauthor{\bsnm{Cooper}, \binits{B.F.}},
\bauthor{\bsnm{Ramakrishnan}, \binits{R.}},
\bauthor{\bsnm{Srivastava}, \binits{U.}},
\bauthor{\bsnm{Silberstein}, \binits{A.}},
\bauthor{\bsnm{Bohannon}, \binits{P.}},
\bauthor{\bsnm{Jacobsen}, \binits{H.-A.}},
\bauthor{\bsnm{Puz}, \binits{N.}},
\bauthor{\bsnm{Weaver}, \binits{D.}},
\bauthor{\bsnm{Yerneni}, \binits{R.}}:
\bctitle{{PNUTS}: {Yahoo!}'s hosted data serving platform}.
In: \bbtitle{Proceedings of the {VLDB} Endowment},
vol. \bseriesno{1},
pp. \bfpage{1277}--\blpage{1288}
(\byear{2008}).
\doiurl{10.14778/1454159.1454167}
\end{bchapter}
\endbibitem

\end{thebibliography}

\end{document}